\documentclass{INTERSPEECH2023}
\usepackage{multirow}

\interspeechcameraready

\title{DuTa-VC: A Duration-aware Typical-to-atypical Voice Conversion Approach with Diffusion Probabilistic Model}
\name{Helin Wang$^1$, Thomas Thebaud$^1$, Jesús Villalba$^{1,2}$, Myra Sydnor$^3$, Becky Lammers$^3$, Najim Dehak$^{1,2}$, 	Laureano Moro-Velazquez$^1$}
\address{ \small
  $^1$Center for Language and Speech Processing, Johns Hopkins University, USA\\
  $^2$Human Language Technology Center of Excellence, Johns Hopkins University, USA \\
  $^3$Department of Physical Medicine and Rehabilitation, Johns Hopkins University School of Medicine, USA}
\email{hwang258@jhu.edu}

\begin{document}

\maketitle
 
\begin{abstract}
We present a novel typical-to-atypical voice conversion approach (DuTa-VC), which (i) can be trained with nonparallel data (ii) first introduces diffusion probabilistic model (iii) preserves the target speaker identity (iv) is aware of the phoneme duration of the target speaker. DuTa-VC consists of three parts: an encoder transforms the source mel-spectrogram into a duration-modified speaker-independent mel-spectrogram, a decoder performs the reverse diffusion to generate the target mel-spectrogram, and a vocoder is applied to reconstruct the waveform. Objective evaluations conducted on the UASpeech show that DuTa-VC is able to capture severity characteristics of dysarthric speech, reserves speaker identity, and significantly improves dysarthric speech recognition as a data augmentation.  Subjective evaluations by two expert speech pathologists validate that DuTa-VC can preserve the severity and type of dysarthria of the target speakers in the synthesized speech.

\end{abstract}

\section{Introduction}
\label{sec:intro}
Voice conversion (VC) is the technique of reproducing a target speaker’s voice while preserving the linguistic content of the utterance pronounced by a source speaker \cite{mohammadi2017overview}.
However, typical-to-atypical (T2A) VC is a largely unexplored frontier, which aims to transform the speech of a typically speaking individual into an atypical one.
Atypical speech in people with dysarthria or dysphonia can be caused by a traumatic injury or a neurological disease \cite{duffy2019motor}. Other etiologies for atypical speech can be related to hearing impairment, intellectual disabilities, and cleft lip or palate, among others. Atypical speech is linked to lower intelligibility through disrupted articulation, phonation, respiration, prosody, resonance, or a combination thereof. 
One important application of T2A VC is data augmentation for atypical automatic speech recognition (ASR) and atypical spoken language understanding (SLU) training, which often suffer from low-resource constraints.
In addition, T2A VC can provide new augmented material to train new caregivers of people with atypical speech and motor disorders, as listener training can improve the listener’s ability to understand atypical speech~\cite{d2006intelligibility}.

For atypical speech recognition, a simple way to do data augmentation is to change the source speaking rate, like speed perturbation \cite{DBLP:conf/interspeech/JinGXYLLM21} and dynamic time warping \cite{harvill2021synthesis}.
However, these methods do not consider other prominent features associated with disordered speech such as articulatory imprecision, breathy and hoarse voice.
Data augmentation methods based on T2A VC have shown promising reductions in word error rates \cite{jin2022personalized,DBLP:conf/interspeech/JinGXYLLM21}.
Among them, frame-wise T2A VC models such as DCGAN \cite{DBLP:conf/interspeech/JinGXYLLM21} and CycleGAN \cite{halpern2021objective} can be used to convert the speech timbre and articulation, but without changing speech rate.
The sequence-to-sequence T2A VC model with Transformer Encoders \cite{harvill2021synthesis,huang2022towards} has the ability to both change the speaking rate and model the atypical characteristics, but it needs parallel data for training, which is hard to collect in real applications, especially for atypical voices.
To achieve nonparallel training, 
autoencoder-based methods like HL-VQ-VAE \cite{illa2021pathological} are employed.
However, these methods change only time-invariant characteristics such as the speaker identity, while preserving time-variant characteristics, such as pronunciation and speaking rate.

To address the above issues, 
inspired by the powerful generative ability of diffusion probabilistic models (DPMs) \cite{DBLP:conf/iclr/ChenZZWNC21,DBLP:conf/iclr/0011SKKEP21},
we propose a nonparallel T2A VC method based on a conditional DPM, called DuTa-VC.
More specifically,
the forward diffusion procedure aims to generate speaker-independent mel-spectrograms from the source mel-spectrograms in order to preserve the linguistic information and remove the speaker information.
This part is jointly trained with a phoneme predictor and a phoneme duration predictor that can be used to modify the phoneme duration to mimic the target voice at the inference stage.
The reverse diffusion procedure with a decoder reconstructs the target mel-spectrograms from the speaker-independent mel-spectrograms conditioned by the speaker information.
Then a vocoder is used to synthesize the waveform.
We conduct experiments on the UASpeech dataset \cite{DBLP:conf/interspeech/KimHPGHWF08},
and the results show that the proposed method can significantly improve the performance of dysarthric speech recognition and our ASR model can achieve comparable results with the state-of-the-art without using parallel data for data augmentation.
Furthermore, subjective evaluations performed by two experienced speech pathologists and objective evaluations employing P-STOI/P-ESTOI metrics and speaker embeddings show that the synthesized voice 
can mimic the intelligibility of dysarthric speech and preserves the target speaker’s characteristics well.

\vspace{-2mm}
\section{Materials and methods}
\begin{figure}[t]
  \centering
  \includegraphics[width=\linewidth]{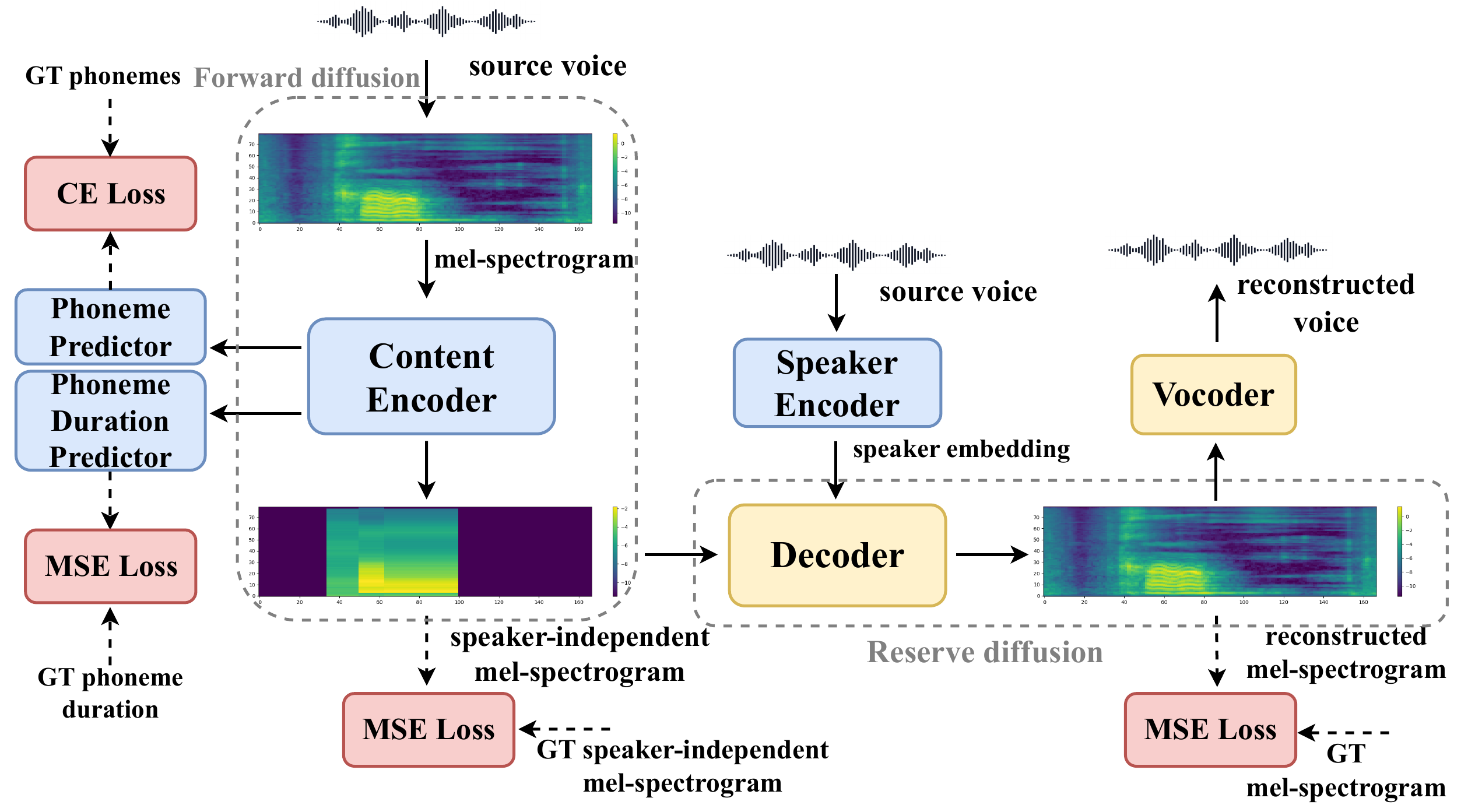}
  \caption{Training Stage of DuTa-VC, with the goal of transforming the source mel-spectrogram into a speaker-independent mel-spectrogram via forward diffusion and converting the speaker-independent mel-spectrogram back to the original one via reverse diffusion. 
    Note that the red boxes denote loss functions, the blue boxes denote modules trained only with typical speech, and the yellow boxes denote modules fine-tuned with atypical speech.  No parallel data is needed for training.}
  \label{fig:training}
    \vspace{-5mm}
\end{figure}

In this section, we introduce the details of our used datasets and the proposed DuTa-VC method.
The basic idea is to transform the source voice into the target atypical voice in the time-frequency domain, \textit{i.e.} using mel-spectrogram instead of raw waveform.
For a conditional DPM with data-dependent prior \cite{popov2021grad,leepriorgrad},
the forward diffusion gradually adds Gaussian noise to data, while the reverse diffusion tries to remove this noise,
and it is trained to minimize the distance between the trajectories of forward and reverse diffusion processes.
The training stage and inference stage of DuTa-VC are shown in Figure~\ref{fig:training} and 
Figure~\ref{fig:inference}, respectively.
Note that 
DuTa-VC is speaker-dependent as we argue that each atypical speaker has different speech intelligibility and specific disrupted articulation and phonation.

\subsection{Datasets}
We use the UASpeech dataset \cite{DBLP:conf/interspeech/KimHPGHWF08}, including word recordings of $15$ dysarthric speakers and $13$ typical control speakers. 
Each dysarthric speaker is categorized into different intelligibility groups: 
very low, low, mid, and high.
The intelligibility was judged by $5$ non-expert American-English native speakers. 
Each speaker has $3$ blocks of $255$ utterances.
Following \cite{jin2022personalized,xiong2019phonetic,xiong2020source},
we use the blocks number $1$ and $3$ for training and the block $2$ for
testing. 
For pre-training the encoder and decoder, described in the next sections, we use typical speech from the large-scale LibriTTS dataset \cite{DBLP:conf/interspeech/ZenDCZWJCW19} which consists of approximately 1100 speakers (10 speakers are held out for validation). All the waveforms are resampled to $22.05$ kHz.

\subsection{Encoder}
The mission of the encoder is to generate a speaker-independent mel-spectrogram (SIMS) representing the same phoneme sequence of the source voice. The content encoder parameterizes the terminal distribution of the forward diffusion, which is only trained with typical speech (LibriTTS).
Following \cite{popovdiffusion},
SIMS is employed as the data-dependent prior, 
which characterizes the phonetic content in time that is independent of the speaker.

To get the ground-truth SIMS for training, we first apply Montreal Forced Aligner \cite{DBLP:conf/interspeech/McAuliffeSM0S17} to the typical training data  to align speech frames with the phonemes extracted from their respective transcriptions.
Next, speaker-independent mel features for each particular phoneme are calculated by aggregating its mel features across the typical speech data.
Each phoneme mel feature of the input mel-spectrograms is then replaced by the average one to get the ground-truth SIMS.
The mean square error (MSE) loss between the output of the content encoder and the ground-truth SIMS is applied.
The encoder is also jointly optimized with a phoneme predictor and a phoneme duration predictor, with a frame-level cross-entropy (CE) loss and MSE loss for training, respectively.
The ground truth phoneme and duration are also obtained by the forced alignment. 

\begin{figure}[t]
  \centering
  \includegraphics[width=\linewidth]{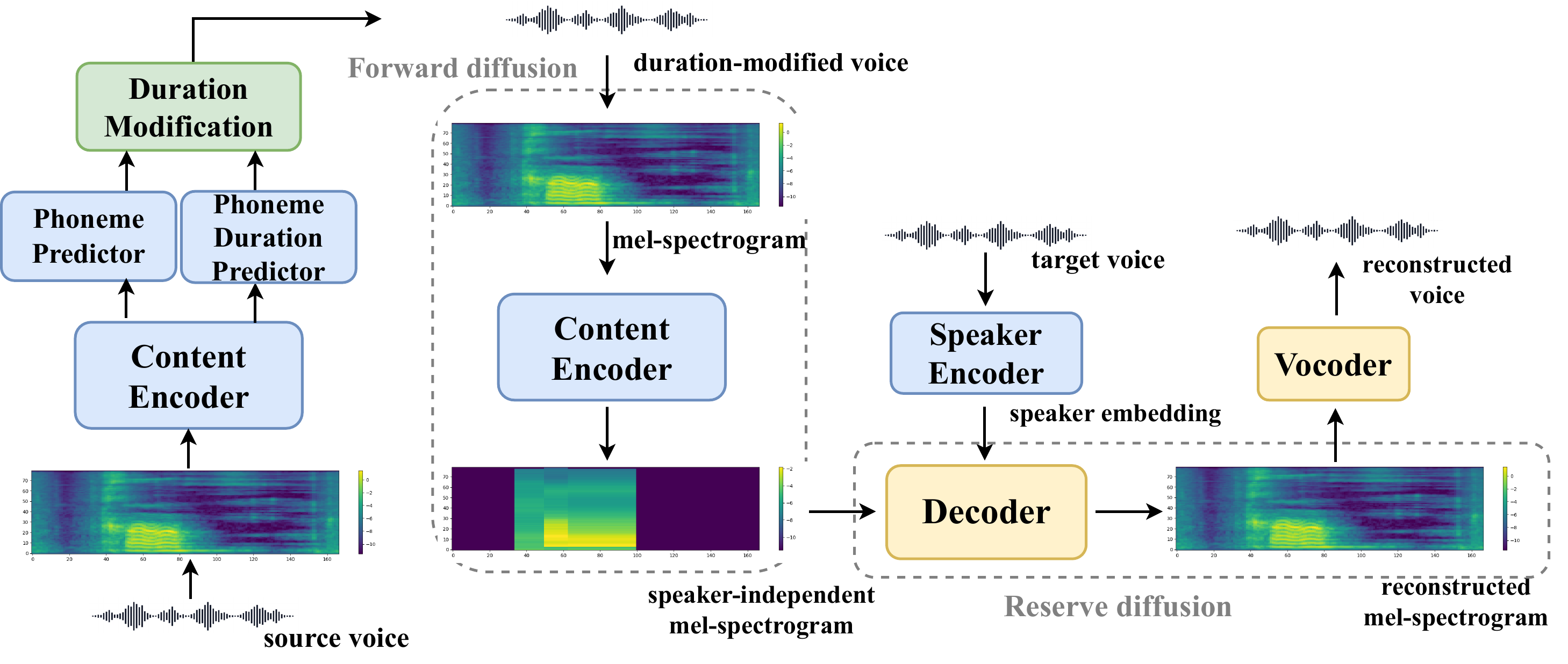}
  \caption{Inference Stage of DuTa-VC. We first predict the phonemes and phoneme duration of the source voice. Next, a duration modification module (the green box) is applied to change the tempo of the source voice. The mel-spectrogram of the duration-modified voice is then used as the input to the encoder. Here, the speaker embedding is obtained from a target atypical voice. }
  \label{fig:inference}
    \vspace{-5mm}
\end{figure}

The goal of introducing the phoneme predictor and the phoneme duration predictor is to achieve phoneme duration modification at inference when generating synthetic atypical speech.
The phoneme predictor simply informs about the estimated phonemes contained in the utterance, whereas the phoneme duration predictor predicts the average phoneme duration in the utterance. Then, for a certain target atypical speaker, we employ the known phoneme duration to stretch or shrink the source audio waveform into a new audio waveform that will have a duration similar to the one that the atypical speaker would have for that phoneme sequence. 
As shown in Figure~\ref{fig:inference}, for a source voice, the phoneme predictor predicts that $N_p$ different phonemes appear in the utterance, and the phoneme duration predictor predicts the mean phoneme duration to be $t_{s}$.
For a target atypical speaker, we calculate the duration of each particular phoneme with forced alignment on the training set, and by averaging the corresponding $N_p$ phoneme duration we can get the mean phoneme duration $t_{t}$.
Tempo perturbation \cite{roberts2019time} is applied to the source voice with a speeding ratio of $t_{t} / t_{s}$ to obtain the duration-modified voice.
The mel-spectrogram of the duration-modified voice is then fed to the content encoder to estimate the SIMS.
Note that we modify the duration of the whole utterance rather than changing each phoneme separately. We will try to change each phoneme duration in future works.

\subsection{Decoder}
The decoder's mission is to generate the target's reconstructed mel-spectrogram given a SIMS and a speaker embedding. The reverse diffusion is parameterized with the decoder, which is pre-trained with typical speech and then fine-tuned with atypical speech.
Following \cite{DBLP:conf/iclr/0011SKKEP21}, we use the forward and reverse diffusions defined by the Stochastic Differential Equations (SDEs):
\begin{align}
d X_t =\frac{1}{2} \left(\bar{X}-X_t\right) \beta_t d t+\sqrt{\beta_t} d \overrightarrow{W_t}, \quad t \in[0, T]
\end{align}
\begin{equation}
\begin{aligned}
d \hat{X}_t= & \left[\frac{1}{2} \left(\bar{X}-X_t\right)-\nabla \log p_t\left(\hat{X}_t\right)\right] \beta_t d t \\
& +\sqrt{\beta_t} d \overleftarrow{W_t} \quad t \in[0, T]
\end{aligned}
\end{equation}
where $\overrightarrow{W}$ and $\overleftarrow{W}$ are two different Wiener processes in $\mathbb{R}^n$, and $n$ is the data dimension. The forward diffusion is to transform data $X_0 \in \mathbb{R}^n$ into the data prior $\bar{X} \in \mathbb{R}^n$ given infinite time horizon $T$,
while the reverse diffusion is to transform $\bar{X}$ back to $X_0$. $\beta_t$ is a non-negative function referred to as noise schedule.

If we consider infinite time horizon, then for any noise schedule $\beta_t$ 
we have $X_t \mid X_0 \stackrel{d}\rightarrow \mathcal{N}(\Bar{X}, I)$ \cite{popov2021grad}.
Thus, if a neural network 
estimates the gradient of the log-density of noisy data $\nabla \log p_t\left(\hat{X}_t\right)$, we can model 
$X_0$
by sampling $X_T$ from $\mathcal{N}(\Bar{X}, I)$ and numerically solving (2) backwards in time.

The described above DPM was introduced in text-to-speech task \cite{popov2021grad} and we further adapt it for the T2A VC task.
We use $\bar{X} = \mathcal{E}\left(X_0\right)$ where $\mathcal{E}\left(\cdot \right)$ is the encoder, i.e. $\bar{X}$ is the SIMS that we want to transform into that of the target voice.
The decoder reconstructs the mel-spectrogram $\hat{X}_t$ = $\mathcal{D}\left(X_t, \bar{X}, t, \varphi_t(Y)\right)$ where 
$\mathcal{D}\left(\cdot \right)$ is the decoder and
$\varphi_t(Y)$ provides the target speaker information. 
$Y$ stands for the source voice at training and the target voice at inference.
Here, we use a pre-trained speaker verification network\footnote{https://github.com/CorentinJ/Real-Time-Voice-Cloning} \cite{DBLP:conf/nips/JiaZWWSRCNPLW18} to extract the speaker embedding $\varphi_t(Y)$.
After generating the mel-spectrogram with the decoder,
the pre-trained universal HiFi-GAN vocoder\footnote{https://github.com/jik876/hifi-gan} \cite{DBLP:conf/nips/KongKB20} is used to reconstruct the audio waveform. We further finetune the vocoder for each atypical speaker.

\vspace{-2mm}
\section{Experiments}

\subsection{Experiment Setups}
For the encoder, we use the same architecture as in Glow-TTS\footnote{https://github.com/jaywalnut310/glow-tts} \cite{DBLP:conf/nips/KimKKY20}, which is composed of a pre-net, $6$ Transformer blocks and a final linear projection layer. 
The phoneme predictor and the phoneme duration predictor borrow the structure from FastSpeech \cite{DBLP:conf/nips/RenRTQZZL19}, which consists of $2$ convolutional layers followed by a projection layer that predicts the probability of phonemes and the logarithm of duration, respectively.
The decoder has the same U-Net\footnote{https://github.com/lucidrains/denoising-diffusion-pytorch} architecture as \cite{ho2020denoising},
which has $4$ feature map resolutions.
Diffusion time $t$ is specified by adding the Transformer sinusoidal position embedding \cite{vaswani2017attention}.
In our experiments, following the settings of \cite{DBLP:conf/nips/KongKB20},
we use $80$-dimensional mel-spectrograms
with 
Short-Time Fourier Transform (STFT) window size of $46.4$ ms and hop size of  $11.6$ ms.
The number of frames for the decoder is $128$ at training,
and we zero-pad mel-spectrograms if the number of frames is not a multiple of $4$ at inference. 
The encoder output $\bar{X}$ is concatenated with $X_t$ as an additional channel.
Following \cite{popovdiffusion}, 
the speaker embedding $\varphi_t(Y)$ is encoded with time positional encoding and passes several linear modules
resulting $128$-dimensional vector which is then broadcast-concatenated as additional $128$ channels.
Thus, the input size of the decoder is $130\times80\times128$ for training.
Our DuTa-VC model has total of $8.5$ million parameters for the encoder and $117.8$ million parameters for the decoder.

When pre-training the encoder and decoder on the LibriTTS, Adam optimizer is employed with initial learning rates $5 \times 10^{-4}$ and $1 \times 10^{-4}$, respectively. Batch sizes are set to $128$ and $32$ with $200$ epochs and $100$ epochs correspondingly.
While for finetuning the decoder on the UASpeech, Adam optimizer is employed with initial learning rate $5 \times 10^{-5}$, batch sizes $32$, and $30$ epochs.
The vocoder is finetuned on the UASpeech with AdamW optimizer \cite{loshchilovdecoupled} of initial learning rate $5 \times 10^{-5}$, batch sizes $128$ and $30$ epochs.
Training segments for reconstruction and the ones used as input to the speaker conditioning network are different random segments sampled from the same training utterances. 
Noise schedule parameters $\beta_t$ are set to $\left[0.05, 20.0\right]$.


 \begin{table}[t]
  \caption{Word error rates (\%) of different ASR systems on the UASpeech. 
    Here, VL, L, M, and H denote respectively very low, low, mid, and high intelligibility groups. 
  }
  \label{tab:1}
    \footnotesize
  \centering
      \vspace{-2mm}
  \begin{tabular}{l|c|c|c|c|c}
    \hline
    \textbf{Method} & \textbf{VL} & \textbf{L} & \textbf{M} & \textbf{H} & 
    \textbf{Avg.} \\
    \hline
    Speech Vision \cite{shahamiri2021speech}&66.3&30.6&40.0&10.5&35.3\\
    SBG+SG \cite{jin2022personalized}&\textbf{61.4}&30.7&22.1&12.8&29.5 \\
    SD-DYS \cite{xiong2019phonetic}&67.8&\textbf{27.6}&26.4&9.7&30.0\\
    SD-Transfer \cite{xiong2020source}&68.2&33.2&22.8&10.4&30.8\\
    SP-DA \cite{geng2020investigation}&66.3&28.6&\textbf{19.9}&9.7&28.5\\
    \hline
    CTRL&95.5&83.9&69.5&29.0&63.8\\
    SYN&77.1&75.9&65.3&28.2&56.7\\
    DYS&72.3&45.0&39.4&12.4&37.3
    \\
    DuTa-VC&65.1&29.7&25.0&9.0&30.1\\
    \textbf{DuTa-VC+}&63.7&27.7&24.1&\textbf{8.5}&\textbf{27.9}
    \\
    \hline
  \end{tabular}
    \vspace{-4mm}
\end{table}

\subsection{Subjective evaluation metrics}
We performed subjective evaluations to measure to which extent the atypical speech traits of the speakers with dysarthria from UASpeech (real targets) were transferred to the synthesized analogous speakers provided by DuTa-VC (synthesized targets). Two expert speech and language pathologists (SLP) with between 10 and 20 years of experience in providing therapy and evaluating the speech of people with dysarthria were asked to perceptually rate the speech of the real targets and the synthesized targets. The SLPs were provided with three recordings from each real target, each containing 20 words from UAspeech. We also provided another three recordings of the synthesized targets (analogous to the real targets, but using different words), containing 20 words each. All the recordings were randomized, and the SLPs did not know how many recordings per speaker were provided or which file corresponded to real or synthetic speech. 
The SLPs rated the overall severity of different atypical speech traits: dysarthria, overall articulation, imprecise consonants, prolonged phonemes, repeated phonemes, irregular articulatory breakdowns, distorted vowels, overall voice quality, harsh voice, hoarse voice, breathiness, strained/strangled voice, stoppages and flutter. The ratings ranged between 0 and 4, where 0 is no severity, and 4 is severe, following the protocol adapted from the Rating Scale for Deviant Speech Characteristics protocol described in chapter three of~\cite{duffy2019motor}, commonly used by SLPs in clinical practice.
Then, for each real target and synthesized analogous target, we averaged the scores of their three respective recordings provided by the two SLPs, separately, and calculated the mean absolute error (MAE), root-mean-square deviation (RMSE) and R-squared (R2) between the ratings of the real and synthesized recordings for each trait. 
Finally, the SLPs were asked to rate the naturalness of all the recordings using a scale between 0 and 4, where 0 meant very synthetic and 4, very natural.



\subsection{Objective evaluation metrics}

\noindent\textbf{Word error rate}:
We used DuTa-VC as a data augmentation method for ASR,
which was implemented using the espnet toolkit\footnote{https://github.com/espnet/espnet/tree/master/egs2/TEMPLATE/asr1}.
This CTC-attention hybrid encoder-decoder model totals $8.7$ million parameters with a vocabulary size of $500$.
The model was trained with Adam optimizer with initial learning rate $1 \times 10^{-3}$, batch size $64$, and $100$ training epochs.
We conducted $5$ ASR systems:
(1) \textbf{CTRL}: the ASR is trained only with the control training data ($6,630$ utterances) and then tested on the real dysarthric test data ($3,825$ utterances).
(2) \textbf{SYN}: the ASR is trained only with the control training data and then tested on the synthesized test data ($3,825$ utterances) to evaluate if the ASR behavior is similar with real and synthetic dysartric speech during inference.
(3) \textbf{DYS}: the ASR is trained with the control training data and the dysarthric training data ($7,650$ utterances), and then tested on the real dysarthric test data.
(4) \textbf{DuTa-VC}: the ASR is trained with the control training data, the dysarthric training data, and the synthesized training data ($99,750$ utterances). For each control training utterance, we convert it targeted to all the $15$ dysarthric speakers as a result of $15$ new training utterances. It is then tested on the real dysarthric test data.
(5) \textbf{DuTa-VC+}: the ASR is trained with the control training data, the dysarthric training data, and the synthesized training data. 
To make the system more robust, we added random Gaussian noises, SpecAugment \cite{park2019specaugment} and changed speed rate with ratio $\{0.9, 1.0, 1.1\}$ for data augmentation purposes during training.
The ASR model is then tested on the real dysarthric test data.


 \noindent\textbf{P-STOI/P-ESTOI}: 
 These metrics are used in our study to quantify the distortion in the time-frequency structure of the signal between the control speech and the dysarthric speech. P-STOI/P-ESTOI
 have been successfully used for the objective evaluation of dysarthric speech \cite{DBLP:conf/icassp/JanbakhshiKB19}. We calculate two values for each metric: one is calculated between the real dysarthric speech and the control speech, the other is calculated between the synthesized speech and the control speech.
 High values indicate low dysarthria severity and good naturalness \cite{huang2022towards}.
 
 
\noindent\textbf{Speaker similarity}: We test the speaker similarity (cosine similarity between speaker embeddings) of source voice, target voice, and generated voice. The goal is to evaluate if DuTa-VC can reproduce the target speaker identity traits.
For each dysarthric speaker, we use $20$ random utterances from each control speaker as source voices to generate synthesized voices.
We compare all the possible control speaker-dysarthric speaker pairs and calculate the mean cosine similarity between them.

 \begin{table}[t]
  \caption{Results of subjective evaluation metrics. $\star$ means that all the values of the real utterances are 0.}
  \label{tab:2}
    \footnotesize
  \centering
  \begin{tabular}{l|c|c|c}
    \hline
    \textbf{Trait} & \textbf{MAE}$\downarrow$ & \textbf{RMSE}$\downarrow$ & \textbf{R$^2$}$\uparrow$ \\
    \hline
    overall dysarthria severity & 0.76 & 0.86& 0.86\\
    overall artic severity &0.66 & 0.77 & 0.91\\
    artic: imprecise consonants &0.67 & 0.78 & 0.87\\
    artic: prolonged phonemes & 0.81 & 0.97 & 0.12\\
    artic: repeated phonemes & 0.23 & 0.48 & 0.85\\
    artic: irregular breakdowns & 0.06 & 0.16 & 0.00$\star$\\
    artic: distorted vowels & 0.64 & 0.85 & 0.55\\
    overall voice quality & 0.39& 0.48 & 0.89\\
    voice: harsh &0.18&0.32 & 0.12\\
    voice: hoarse/wet & 0.14&0.27 & 0.37\\
    voice: breathy &0.36&0.54&0.53\\
    voice: strained/strangled & 0.46&0.51&0.85\\
    voice: stoppages &0.04&0.09&0.19\\
    voice: flutter &0.09&0.14&0.00$\star$\\
    \hline
  \end{tabular}
  \vspace{-4mm}
\end{table}
 
\subsection{Results}
The performance of different ASR systems on UASpeech is shown in Table~\ref{tab:1}. We compare  our results with those reported in five previous studies employing UASpeech for ASR, introduced in Section \ref{sec:intro}.
When no atypical speech is used for training,
CTRL yields the average WER of $63.8\%$. 
SYN provides an average WER of $56.7\%$ that is close to the results of CTRL, which suggests that the ASR system shows similar sensitivity to the real and synthesized data.
The proposed DuTa-VC data augmentation significantly improves the performance of ASR compared to DYS.
In addition, DuTa-VC can work well with other data augmentation methods and DuTa-VC+ provides the best average WER of $27.9\%$.
Compared to previous methods, DuTa-VC+ leads to the best overall results, as it performs well across several intelligibility degrees, whereas previous methods tend to work well in only one.
For instance, the GAN-based method proposed in \cite{jin2022personalized} yields the best result in the VL group, but it underperforms in other groups and needs parallel data for training. 
Speed perturbation and speaker adaptive training were used in \cite{geng2020investigation} and provided the best results in the M group. However, this approach does not perform very well in other severities.
Phoneme-based speech tempo adjustments were used in \cite{xiong2019phonetic,xiong2020source}, but they need a manual data selection.
In contrast, DuTa-VC+ performs well across all the groups and achieves similar results to the best WER per group in previous studies, yielding the best overall WER, while not requiring parallel data.

 \begin{table}[t]
  \caption{Results of P-STOI and P-ESTOI tests (ranging 0-1). 
  }
  \label{tab:3}
    \footnotesize
  \centering
      \vspace{-2mm}
  \begin{tabular}{ l|c|c|c|c}
    \hline
    \textbf{Metric} & \textbf{VL} & \textbf{L} & \textbf{M} & \textbf{H} \\
    \hline
    P-STOI (Real) &0.42&0.50&0.58&0.69\\
    P-STOI (Synthesized) &0.49&0.54&0.62&0.71\\
    P-ESTOI (Real) &0.08&0.15&0.19&0.34\\
    P-ESTOI (Synthesized) &0.11&0.18&0.20&0.35\\
    \hline
  \end{tabular}
  \vspace{-1mm}
\end{table}


 \begin{table}[t]
  \caption{Results of the speaker similarity experiments. S, T, and G denote source voices, target voices and generated voices.}
  \label{tab:4}
    \footnotesize
  \centering
      \vspace{-2mm}
  \begin{tabular}{ c|c|c}
    \hline
     \textbf{S with T} & \textbf{S with G} & \textbf{T with G}\\
    \hline
    78.6\%&77.9\%&94.6\%
    \\
    \hline
  \end{tabular}
  \vspace{-4mm}
\end{table}

Table~\ref{tab:2} shows the results of subjective evaluation metrics\footnote{The perceptual evaluations of the UASpeech dysarthric speakers and audio demos are shared to allow further studies: https://wanghelin1997.github.io/DuTa-VC-Demo/}.
All the average MAE in perceptual evaluation between real and synthetic speech are less than $0.81$, which shows well-matched severity of dysarthria and articulation between real and synthetic targets. Among them, DuTa-VC can model the articulatory breakdowns, harsh voice, hoarse/wet voice, stoppages, and flutter well (MAE below $0.25$). 
Regarding naturalness, synthetic targets were rated $1.15$ points lower on average, which means that these were perceived as more synthetic than the real recordings. Since each file contained 20 words, a glitch in a single word could lead the SLP evaluators to rate the whole recording with lower naturalness. In this sense, word-wise or utterance-wise comparisons could have led to a higher naturalness score of the synthetic speech.

Table~\ref{tab:3} summarizes the results of P-STOI and P-ESTOI analyses.
With the speech intelligibility increasing (from the VL group to the H group), the P-STOI and P-ESTOI get higher values which means the severity decreases.
In summary,
the synthesized dysarthric voices by DuTa-VC show similar dysarthria severity and naturalness with real dysarthric voices.
On the other hand, Table~\ref{tab:4} shows the identity preservation ability of DuTa-VC.
The results show that the synthetic generated voices have quite a high similarity with the real target voices while the generated voices and the target voices both have much less similarity with the source voices (non-target).

\section{Conclusions and future work}
In this paper,
we proposed a novel duration-aware typical-to-atypical voice conversion approach
\footnote{The source code and trained models will be released at https://github.com/WangHelin1997/DuTa-VC.} 
based on the diffusion probabilistic model.
 Experimental results on UASpeech show the method can significantly improve the performance of ASR, 
 capture severity characteristics of atypical speech, and preserve speaker identity well.
In future works, we will
(1) adapt our method to sentence utterances instead of words, using a new dataset we have started collecting.
(2) explore new, more accurate ways to modify the duration for each phoneme separately.
(3) use the augmented data to improve spoken language understanding in dysarthric speech.

{
\tiny
\bibliographystyle{IEEEtran}
\bibliography{mybib}}

\end{document}